\documentclass[10pt,a4paper,twoside,twocolumn,english,aps,pra,superscriptaddress,twocolumn,showpacs,amsmath,amssymb,floatfix]{revtex4}

\usepackage[T1]{fontenc}
\usepackage[latin9]{inputenc}
\usepackage{amsmath}
\usepackage{amssymb}
\usepackage{graphicx}
\usepackage{esint}

\makeatletter

\@ifundefined{textcolor}{}
{%
 \definecolor{BLACK}{gray}{0}
 \definecolor{WHITE}{gray}{1}
 \definecolor{RED}{rgb}{1,0,0}
 \definecolor{GREEN}{rgb}{0,1,0}
 \definecolor{BLUE}{rgb}{0,0,1}
 \definecolor{CYAN}{cmyk}{1,0,0,0}
 \definecolor{MAGENTA}{cmyk}{0,1,0,0}
 \definecolor{YELLOW}{cmyk}{0,0,1,0}
}

\makeatother

\usepackage{babel}
\begin{document}

\title{Bogoliubov Theory of Dipolar Bose Gas in Weak Random Potential}

\author{Mahmoud Ghabour}

\affiliation{Insitute f\"ur Theoretische Physik, Freie Universit\"at Berlin,
Arnimallee 14, 14195 Berlin, Germany}

\author{Axel Pelster}

\affiliation{Physics Department and Research Center OPTIMAS, Technical University
of Kaiserslautern, 67663 Kaiserslautern, Germany}

\affiliation{Hanse-Wissenschaftskolleg, Lehmkuhlenbusch 4, D-27733 Delmenhorst,
Germany }
\begin{abstract}
We consider a dilute homogeneous Bose gas with both an isotropic short-range
contact interaction and an anisotropic long-range dipole-dipole interaction
in a weak random potential at low temperature in three dimensions.
Within the realm of Bogoliubov theory we analyze how both condensate
and superfluid are depleted due to quantum and thermal fluctuations
as well as disorder fluctuations.
\end{abstract}


\maketitle

\section{Introduction}

Since the first observation of Bose-Einstein condensation (BEC) in
1995 for alkali atomic vapors of $^{87}$Rb and $^{23}$Na atoms both
theoretical and experimental research has accelerated to study this
newly discovered macroscopic quantum phenomenon, where a fraction
of these bosons occupies the same quantum mechanical ground state.
Achieving BEC in experiment for alkali atoms was only possible due
to the discovery of efficient cooling and trapping techniques which
became available by the end of the twentieth century \cite{key-3,key-1,key-2}.
With more cooling advances, research interest has not only increased
in the field of ultra-cold quantum gases with isotropic short-range
contact interaction, which represents the effective interaction between
atoms and is dominated by the s-wave scattering length. Also the anisotropic
long-range dipole-dipole interaction has been made accessible to detailed
study by the formation of a BEC in a dipolar quantum gas of $^{52}$Cr
atoms \cite{pfau,pfau-3}. Further atomic BECs with magnetic dipole-dipole
interaction followed soon with $^{164}$Dy \cite{lev} and $^{168}$Er
\cite{grimm} atoms. Strong electric dipole-dipole interactions were
realized by a stimulated Raman adiabatic passage (STIRAP) experiment,
which allowed the creation of $^{40}$K $^{87}$Rb molecules in the
rovibrational ground state \cite{ospelkhaus,ospelkhaus1}.

The research field of ultra-cold quantum gases allows today to analyze
many condensed matter problems in the spirit of Feynman's quantum
simulator \cite{Fenman}, as all ingredients of the quantum many-body
Hamiltonian can be experimentally tuned with an unprecedented precision.
The kinetic energy can be controlled with spin-orbit coupling \cite{kinetic,kinetic1},
the potential energy can be harmonic \cite{harmonic} box-like \cite{hadzib}
or even anharmonic \cite{aharmonic} and also the strength of contact
as well as dipolar interaction are tunable \cite{feshbach,pfau-1,pfau-2}.
In addition, in order to make quantum gas simulators even more realistic,
various experimental and theoretical methods for controlling the effect
of disorder have been designed. Using superfluid helium in vycor glass,
which represents some kind of porous media, the random distribution
of pores represents a random environment \cite{key-4}. Laser speckles
are produced by focusing the laser beam on a glass plate, where the
resulting random interference pattern is reflected to a BEC cell \cite{key-5,key-7,key-6,aspect}.
Wire traps represent magnetic traps on atomic chips where the roughness
and the imperfection of the wire surface generates a disorder potential
\cite{key-8,key-9}. Another possibility to create a random potential
is to trap one species of atoms randomly in a deep optical lattice,
which serves as frozen scatterers for a second atomic species \cite{key-10,Schneble}.
In addition also, incommensurable lattices provide a useful random
environment \cite{incommen1,incommen2,incommen3}.

In order to study the properties of interacting bosons in such a random potential
Huang and Meng proposed a Bogoliubov theory, which was applied to
the case of superfluid helium in porous media \cite{key-11}, and
extended later by others \cite{giorgini,lopatin,key-12,graham1}. For a delta-correlated
disorder it was found that both a condensate and a superfluid depletion
occurs due to the localization of bosons in the respective minima
of the external random potential which is present even at zero temperature.
A generalization to the corresponding situation, where the disorder
correlation function falls off with a characteristic correlation length
as, for instance, a Gauss function \cite{key-14,pelster1}, laser
speckles \cite{pelster2} or a Lorentzian \cite{pelster3} is straightforward.

In this paper we extend the Bogoliubov theory of Huang and Meng \cite{key-11}
to a dipolar Bose gas at finite temperature. With this we go beyond
the zero-temperature mean-field approach where the Gross-Pitaevskii
equation is solved perturbativly with respect to the random potential
\cite{pelster1,pelster3}. This extension to the finite-temperature
regime allows us to study in detail how the anisotropy of superfluidity
can be tuned, a phenomenon which also occurs at zero-temperature \cite{pelster1,pelster3},
but turns out to be enhanced by the thermal fluctuations. We observe in addition
that contact and dipolar interaction have different effects upon quantum,
thermal, and disorder depletion of condensate and superfluid.

The paper is organized as follows. In Section. II we revisit the Bogoliubov
theory of the homogeneous dirty boson problem. With this, we determine
for a general two-particle interaction and a general disorder correlation
for different observables the beyond mean-field corrections which
stem from quantum, thermal, and disorder fluctuations. In Section. III
we specialize our treatment for a dipolar Bose gas and a delta-correlated
disorder in the zero-temperature and the thermodynamic limit. In particular,
we investigate the particle and condensate density as well as the
inner energy. In Section. IV we extend the Bogoliubov theory in order
to derive the superfluid depletion which turns out to be only due
to the external random potential and the thermal excitation. In Section.
V we consider how finite-temperature effects on the condensate depletion
as well as the normal fluid component depend on the respective strength
of contact and dipolar interaction.

\section{Bogoliubov Theory}

A three-dimensional ultra-cold dipolar Bose gas in a weak random potential
is modeled by the grand-canonical Hamiltonian
\begin{alignat}{1}
\mathcal{\hat{K}}= & \int d^{3}\mathbf{x}\,\hat{\psi}^{\dagger}(\mathbf{x})\left[\frac{-\hbar^{2}\nabla^{2}}{2m}-\mu+U(\mathbf{x})\right]\hat{\psi}(\mathbf{x})\label{eq:1}\\
 & +\frac{1}{2}\int d^{3}\mathbf{x}\int d^{3}\mathbf{x}^{\prime}\hat{\psi}^{\dagger}(\mathbf{x})\hat{\psi}^{\dagger}(\mathbf{x^{\prime}})V(\mathbf{x},\mathbf{x^{\prime}})\hat{\psi}(\mathbf{x^{\prime}})\hat{\psi}(\mathbf{x}),\nonumber 
\end{alignat}
where $\hat{\psi}(\mathbf{x})$ and $\hat{\psi}^{\dagger}(\mathbf{x})$
are the usual field operators for Bose particles of mass $m$, which
satisfy the following commutation relations

\begin{alignat}{1}
\Bigl[\hat{\psi}(\mathbf{x}),\hat{\psi}^{\dagger}(\mathbf{x^{\prime}})\Bigr]= & \delta(\mathbf{x}-\mathbf{x^{\prime}}),\nonumber \\
\Bigl[\hat{\psi}(\mathbf{x}),\hat{\psi}(\mathbf{x^{\prime}})\Bigr]= & \Bigl[\hat{\psi}^{\dagger}(\mathbf{x}),\hat{\psi}^{\dagger}(\mathbf{x^{\prime}})\Bigr]=0.\label{eq:10}
\end{alignat}
Here $\mu$ denotes the chemical potential and $U(\mathbf{x})$ represents
the random potential. Irrespective of the physical origin of the disorder
potential, we assume that the average over the random potential vanishes
and that it has some kind of correlation i.e. $\langle U(\mathbf{x})\rangle=0$
and $\langle U(\mathbf{x})U(\mathbf{x}^{\prime})\rangle=R(\mathbf{x}-\mathbf{x}^{\prime})$.
Furthermore, the two-particle interaction potential for a dipolar
Bose gas consists of two different parts: $V(\mathbf{x},\mathbf{x^{\prime}})=V_{\delta}(\mathbf{x}-\mathbf{x^{\prime}})+V_{\mathrm{dd}}(\mathbf{x}-\mathbf{x^{\prime}})$.
On the one hand the short-range isotropic contact interaction is given
by $V_{\delta}(\mathbf{x})=g\delta(\mathbf{x})$, where $g=4\pi a\hbar^{2}/m$
represents its strength with the s-wave scattering length $a$. From
now on we assume a positive $a$, i.e. a repulsive contact interaction,
which depletes the particles from the ground state due to quantum
and thermal fluctuations and also forbids them from being localized
in the minima of the external random potential $U(\mathbf{x})$. Thus,
unlike the case of free fermions, where Pauli blocking prevents the
particles from localizing in a single orbital, bosons require a repulsive
interaction to prevent them from collapsing into the respective minima
of the external random potential. On the other hand the long-range
anisotropic polarized dipolar part is written in real space for dipoles
aligned along $z$-axis direction according to 
\begin{alignat}{1}
V_{\mathrm{dd}}(\mathbf{x})=\frac{C_{\mathrm{dd}}}{4\pi}\:\frac{(x^{2}+y^{2}+z^{2})-3z^{2}}{(x^{2}+y^{2}+z^{2})^{5/2}} & ,\label{eq:3}
\end{alignat}
where $C_{\mathrm{dd}}$ represents the dipolar interaction strength
due to magnetic or electric dipole moments. In the first case we have
$C_{\mathrm{dd}}=\mu_{0}d_{\mathrm{m}}^{2}$, with the magnetic dipole
moment $d_{\mathrm{m}}$ and the magnetic permeability of vacuum $\mu_{0}$,
and in the latter case we have $C_{\mathrm{dd}}=d_{\mathrm{e}}^{2}/\epsilon_{0}$,
with the electric dipole moment $d_{\mathrm{e}}$ and the vacuum dielectric
constant $\epsilon_{0}$. Note that, due to the anisotropic character
of the dipolar interaction, many static and dynamic properties of
a dipolar Bose gas become tunable \cite{pfau,pfau-3,glaum,glaum1},
consequences for a random environment have only recently be explored
\cite{pelster1,pelster3}.

With the help of a Fourier transformation we can rewrite the Hamiltonian
(\ref{eq:1}) in momentum space according to 
\begin{alignat}{1}
\mathcal{\hat{K}}= & \sum_{\mathbf{k}}\left(\frac{\hbar^{2}\mathbf{k}^{2}}{2m}-\mu\right)\hat{a}_{\mathbf{k}}^{\dagger}\hat{a}_{\mathbf{k}}+\frac{1}{v}\sum_{\mathbf{p},\mathbf{k}}U_{\mathbf{p}-\mathbf{k}}\;\hat{a}_{\mathbf{p}}^{\dagger}\hat{a}_{\mathbf{k}}\nonumber \\
 & +\frac{1}{2v}\sum_{\mathbf{p},\mathbf{k},\mathbf{q}}V_{\mathbf{q}}\:\hat{a}_{\mathbf{k}+\mathbf{q}}^{\dagger}\hat{a}_{\mathbf{p}-\mathbf{q}}^{\dagger}\hat{a}_{\mathbf{p}}\hat{a}_{\mathbf{k}},\label{eq:4}
\end{alignat}
where $v$ denotes the volume and the interaction potential in momentum
space is given by \cite{pfau-1}

\begin{alignat}{1}
V_{\mathbf{q}}= & g+\frac{C_{\mathrm{dd}}}{3}\left(3\cos^{2}\theta-1\right).\label{eq:92}
\end{alignat}
Here $\theta$ denotes the angle between the polarization direction,
which is here along the $z$-axis, and the wave vector $\mathbf{q}$.
Note that (\ref{eq:92}) is not continuous at $\mathbf{q}=\mathbf{0}$,
as the limit $\mathbf{q}\rightarrow\mathbf{0}$ is direction dependent.
This is the origin for various anisotropic properties, which are characteristic
for dipolar Bose gases \cite{pelster1,pelster3,key-15,lima}. The
operators $\hat{a}_{\mathbf{k}}$ and $\hat{a}_{\mathbf{k}}^{\dagger}$
are the annihilation and creation operators in Fourier space, respectively,
which turn out to satisfy the bosonic commutation relations

\begin{alignat}{1}
\Bigl[\hat{a}_{\mathbf{k}},\hat{a}_{\mathbf{\mathbf{k}^{\prime}}}^{\dagger}\Bigr]=\delta_{\mathbf{k}\mathbf{k}^{\prime}},\;\Bigl[\hat{a}_{\mathbf{k}},\hat{a}_{\mathbf{\mathbf{k}^{\prime}}}\Bigr]=\Bigl[\hat{a}_{\mathbf{k}}^{\dagger},\hat{a}_{\mathbf{\mathbf{k}^{\prime}}}^{\dagger}\Bigr]=0 & .\label{eq:11}
\end{alignat}
Near absolute zero temperature the number of the particles $N_{\mathbf{0}}$
in the ground state $|\Phi_{\mathbf{0}}\rangle$ becomes macroscopically
large. In this case we have due to (\ref{eq:11}) $\hat{a}_{\mathbf{0}}|\Phi_{\mathbf{0}}\rangle\thickapprox\sqrt{N_{\mathbf{0}}}|\Phi_{\mathbf{0}}\rangle$
and $\hat{a}_{\mathbf{0}}^{\dagger}|\Phi_{\mathbf{0}}\rangle\thickapprox\sqrt{N_{\mathbf{0}}}|\Phi_{\mathbf{0}}\rangle$,
so the operators $\hat{a}_{\mathbf{0}}$ and $\hat{a}_{\mathbf{0}}^{\dagger}$
approximately commute with each other. Thus we can apply the Bogoliubov
prescription \cite{bogoliubov} and replace the creation and annihilation
operators of the ground state $\mathbf{k}=\mathbf{0}$ by a c-number,
i.e. $\hat{a}_{\mathbf{0}}=\hat{a}_{\mathbf{0}}^{\dagger}=\sqrt{N_{\mathbf{0}}}$.
As a consequence, we have to decompose the respective momentum summations
in the Hamiltonian (\ref{eq:4}) into their ground state $\mathbf{k}=\mathbf{0}$
and excited states $\mathbf{k}\neq\mathbf{0}$ contributions. By doing
so, we perform the following two physical approximations. At first,
we ignore terms which contain creation and annihilation operators
of the excited states $\mathbf{k}\neq\mathbf{0}$, which are of third
and fourth order, as they represent higher-order interactions of the
particles out of the condensate. Such an approximation is justified
in the case of weakly interacting systems. Secondly, we assume for
weak enough disorder that disorder fluctuations decouple in lowest
order. Therefore, we ignore the terms $U_{\mathbf{p}-\mathbf{k}}\;\hat{a}_{\mathbf{p}}^{\dagger}\hat{a}_{\mathbf{k}}$
with both $\mathbf{k}\neq\mathbf{0}$ and $\mathbf{p}\neq\mathbf{0}$.
Note that these two physical approximations imply that the excited
states are only rarely occupied, i.e. $N-N_{\mathbf{0}}\ll N$. With
this we get the simplified Hamiltonian

\begin{alignat}{1}
\mathcal{\hat{K}}^{\prime}= & \left(-\mu+\frac{1}{v}U_{0}\right)N_{\mathbf{0}}+\frac{1}{2v}V_{\mathbf{0}}N_{\mathbf{0}}^{2}\nonumber \\
 & +\frac{1}{2}\sum_{\mathbf{k}}^{\prime}\left(\frac{\hbar^{2}\mathbf{k}^{2}}{2m}-\mu\right)(\hat{a}_{\mathbf{k}}^{\dagger}\hat{a}_{\mathbf{k}}+\hat{a}_{-\mathbf{k}}^{\dagger}\hat{a}_{-\mathbf{k}})\nonumber \\
 & +\frac{1}{v}\sqrt{N_{\mathbf{0}}}\sum_{\mathbf{k}}^{\prime}U_{\mathbf{k},\mathbf{0}}\;(\hat{a}_{\mathbf{k}}^{\dagger}+\hat{a}_{-\mathbf{k}})\nonumber \\
 & +\frac{1}{2v}N_{\mathbf{0}}\sum_{\mathbf{k}}^{\prime}(V_{\mathbf{0}}+V_{\mathbf{k}})\:(\hat{a}_{\mathbf{k}}^{\dagger}\hat{a}_{\mathbf{k}}+\hat{a}_{-\mathbf{k}}^{\dagger}\hat{a}_{-\mathbf{k}})\nonumber \\
 & +\frac{1}{2v}N_{\mathbf{0}}\sum_{\mathbf{k}}^{\prime}V_{\mathbf{k}}\:(\hat{a}_{\mathbf{k}}^{\dagger}\hat{a}_{-\mathbf{k}}^{\dagger}+\hat{a}_{\mathbf{k}}\hat{a}_{-\mathbf{k}}).\label{eq:62-1}
\end{alignat}
Here the prime over the summation symbol indicates that the ground
state $\mathbf{k}=\mathbf{0}$ is excluded. The appearance of off-diagonal
terms $\hat{a}_{\mathbf{k}}\hat{a}_{-\mathbf{k}}$ and $\hat{a}_{\mathbf{k}}^{\dagger}\hat{a}_{-\mathbf{k}}^{\dagger}$
with $\mathbf{k}\neq\mathbf{0}$ in equation (\ref{eq:62-1}) indicates
that the state $\mathbf{k}=\mathbf{0}$ is no longer the ground state
of the interacting system. In order to determine this new mean-field
ground state we have to diagonalize the Hamiltonian (\ref{eq:62-1}).
To this end we follow Ref.~\cite{key-11} and use the inhomogeneous
Bogoliubov transformation

\begin{alignat}{1}
\hat{a}_{\mathbf{k}}= & u_{\mathbf{k}}\hat{\alpha}_{\mathbf{k}}-v_{\mathbf{k}}\hat{\alpha}_{-\mathbf{k}}^{\dagger}-z_{\mathbf{k}},\nonumber \\
\hat{a}_{\mathbf{k}}^{\dagger}= & u_{\mathbf{k}}\hat{\alpha}_{\mathbf{k}}^{\dagger}-v_{\mathbf{k}}\hat{\alpha}_{-\mathbf{k}}-z_{\mathbf{k}}^{*}.\label{eq:63}
\end{alignat}
Note that the Bogoliubov amplitudes $u_{\mathbf{k}}$ and $v_{\mathbf{k}}$ can be chosen to be real without
loss of generality. Furthermore, we impose upon the new operators
$\hat{\alpha}_{\mathbf{k}},\:\hat{\alpha}_{\mathbf{k}}^{\dagger}$
that they also satisfy bosonic commutation relations

\begin{alignat}{1}
\Bigl[\hat{\alpha}_{\mathbf{k}},\hat{\alpha}_{\mathbf{k}^{\prime}}^{\dagger}\Bigr]=\delta_{\mathbf{k}\mathbf{k}^{\prime}},\;\Bigl[\hat{\alpha}_{\mathbf{k}},\hat{\alpha}_{\mathbf{k}^{\prime}}\Bigr]=\Bigl[\hat{\alpha}_{\mathbf{k}}^{\dagger},\hat{\alpha}_{\mathbf{k}^{\prime}}^{\dagger}\Bigr]=0 & .\label{eq:50}
\end{alignat}
Inserting (\ref{eq:63}) in the simplified Hamiltonian (\ref{eq:62-1})
we obtain via diagonalization the following results for the Bogoliubov
parameters
\begin{alignat}{1}
u_{\mathbf{k}}^{2}= & \frac{1}{2}\left[\frac{\frac{\hbar^{2}\mathbf{k}^{2}}{2m}-\mu+n_{\mathbf{0}}(V_{\mathbf{0}}+V_{\mathbf{k}})}{E_{\mathbf{k}}}+1\right],\nonumber \\
v_{\mathbf{k}}^{2}= & \frac{1}{2}\left[\frac{\frac{\hbar^{2}\mathbf{k}^{2}}{2m}-\mu+n_{\mathbf{0}}(V_{\mathbf{0}}+V_{\mathbf{k}})}{E_{\mathbf{k}}}-1\right],
\end{alignat}
and the translations
\begin{alignat}{1}
z_{\mathbf{k}}= & \frac{\frac{1}{v}\sqrt{N_{\mathbf{0}}}U_{\mathbf{k}}\left(\frac{\hbar^{2}\mathbf{k}^{2}}{2m}-\mu+n_{\mathbf{0}}V_{\mathbf{0}}\right)}{E_{\mathbf{k}}^{2}}.\label{eq:9}
\end{alignat}
Here we have introduced for brevity the condensate density $n_{\mathbf{0}}=N_{\mathbf{0}}/v$
and the quasi-particle dispersion
\begin{alignat}{1}
E_{\mathbf{k}}=\sqrt{\left[\frac{\hbar^{2}\mathbf{k}^{2}}{2m}-\mu+n_{\mathbf{0}}\left(V_{\mathbf{0}}+V_{\mathbf{k}}\right)\right]^{2}-\left(n_{\mathbf{0}}V_{\mathbf{k}}\right)^{2}} & .\label{eq:10-1}
\end{alignat}
Note that the diagonalized Hamiltonian changes with each realization
of the disorder potential $U(\mathbf{x})$. Therefore, we get the
final Hamiltonian of the dirty dipolar Bose system by performing the
disorder ensemble average
\begin{alignat}{1}
\left\langle \mathcal{\hat{K}}^{\prime}\right\rangle = & v\left(-\mu n_{\mathbf{0}}+\frac{1}{2}V_{\mathbf{0}}n_{\mathbf{0}}^{2}\right)\nonumber \\
 & +\frac{1}{2}\sum_{\mathbf{k}}^{\prime}\left\{ E_{\mathbf{k}}-\left[\frac{\hbar^{2}\mathbf{k}^{2}}{2m}-\mu+n_{\mathbf{0}}(V_{\mathbf{0}}+V_{\mathbf{k}})\right]\right\} \nonumber \\
 & +\frac{1}{2}\sum_{\mathbf{k}}^{\prime}E_{\mathbf{k}}(\hat{\alpha}_{\mathbf{k}}^{\dagger}\hat{\alpha}_{\mathbf{k}}+\hat{\alpha}_{-\mathbf{k}}^{\dagger}\hat{\alpha}_{-\mathbf{k}})\nonumber \\
 & -\sum_{\mathbf{k}}^{\prime}\frac{n_{\mathbf{0}}R_{\mathbf{k}}}{E_{\mathbf{k}}^{2}}\left(\frac{\hbar^{2}\mathbf{k}^{2}}{2m}-\mu+n_{\mathbf{0}}V_{\mathbf{0}}\right).\label{eq:12}
\end{alignat}
With this it is straightforward to determine the corresponding grand-canonical
potential $\Omega{}_{\mathrm{eff}}=-\beta^{-1}\,\ln\,\mathcal{Z}_{G}$,
where $\mathcal{Z}_{G}=\mathrm{Tr}\, e^{-\beta\left\langle \mathcal{\hat{K}}^{\prime}\right\rangle }$
denotes the grand-canonical partition function and $\beta=1/\left(k_{\mathrm{B}}T\right)$
is the reciprocal temperature:
\begin{alignat}{1}
\Omega{}_{\mathrm{eff}}=v & \left(-\mu n_{\mathbf{0}}+\frac{1}{2}V_{\mathbf{0}}n_{\mathbf{0}}^{2}\right)\nonumber \\
 & +\frac{1}{2}\sum_{\mathbf{k}}^{\prime}\left\{ E_{\mathbf{k}}-\left[\frac{\hbar^{2}\mathbf{k}^{2}}{2m}-\mu+n_{\mathbf{0}}(V_{\mathbf{0}}+V_{\mathbf{k}})\right]\right\} \nonumber \\
 & +\sum_{\mathbf{k}}^{\prime}\frac{1}{\beta}\mathrm{ln}\left(1-e^{-\beta E_{\mathbf{k}}}\right)\nonumber \\
 & -\sum_{\mathbf{k}}^{\prime}\frac{n_{\mathbf{0}}R_{\mathbf{k}}}{E_{\mathbf{k}}^{2}}\left(\frac{\hbar^{2}\mathbf{k}^{2}}{2m}-\mu+n_{\mathbf{0}}V_{\mathbf{0}}\right).\label{eq:13}
\end{alignat}
Extremizing equation (\ref{eq:13}) with respect to the condensate
density $n_{\mathbf{0}}$ for fixed chemical potential $\mu$ we find
up to first order in quantum, thermal, and disorder fluctuations
\begin{alignat}{1}
n_{\mathbf{0}}= & \frac{\mu}{V_{\mathbf{0}}}-\frac{1}{2v}\sum_{\mathbf{k}}^{\prime}\Biggl\{\frac{\left(\frac{\hbar^{2}\mathbf{k}^{2}}{2m}-\mu\right)(V_{\mathbf{0}}+V_{\mathbf{k}})}{V_{\mathbf{0}}E_{\mathbf{k}}}\label{eq:14}\\
 & \qquad\frac{+\frac{\mu}{V_{\mathbf{0}}}(V_{\mathbf{0}}^{2}+2V_{\mathbf{0}}V_{\mathbf{k}})-E_{\mathbf{k}}(V_{\mathbf{0}}+V_{\mathbf{k}})}{V_{\mathbf{0}}E_{\mathbf{k}}}\Biggr\}\nonumber \\
 & -\frac{1}{v}\sum_{\mathbf{k}}^{\prime}\frac{\left(\frac{\hbar^{2}\mathbf{k}^{2}}{2m}-\mu\right)(V_{\mathbf{0}}+V_{\mathbf{k}})+\frac{\mu}{V_{\mathbf{0}}}(V_{\mathbf{0}}^{2}+2V_{\mathbf{0}}V_{\mathbf{k}})}{\left(e^{\beta E_{\mathbf{k}}}-1\right)V_{\mathbf{0}}E_{\mathbf{k}}}\nonumber \\
 & +\frac{1}{v}\sum_{\mathbf{k}}^{\prime}\frac{R_{\mathbf{k}}}{E_{\mathbf{k}}^{4}}\left\{ \frac{1}{V_{\mathbf{0}}}\left(\frac{\hbar^{2}\mathbf{k}^{2}}{2m}\right)^{3}-\frac{\mu}{V_{\mathbf{0}}}\left(\frac{\hbar^{2}\mathbf{k}^{2}}{2m}\right)^{2}\right\} ,\nonumber 
\end{alignat}
where the quasi-particle dispersion (\ref{eq:10-1}) now reads 
\begin{alignat}{1}
E_{\mathbf{k}}=\sqrt{\left(\frac{\hbar^{2}\mathbf{k}^{2}}{2m}\right)^{2}+\mu\frac{\hbar^{2}\mathbf{k}^{2}}{m}\frac{V_{\mathbf{k}}}{V_{\mathbf{0}}}} & .\label{eq:16}
\end{alignat}
Note that the zeroth order $n_{\mathbf{0}}=\mu/V_{\mathbf{0}}$ in
(\ref{eq:14}) represents the mean-field result and corresponds to
the Hugenholtz-Pines relation \cite{pines}. It makes the quasi-particle
dispersion (\ref{eq:10-1}) according to (\ref{eq:16}) linear and
gapless (massless) in the long-wavelength limit $\mathbf{k}\rightarrow\mathbf{0}$,
in accordance with the Nambu-Goldstone theorem \cite{nambu,goldstone}.

Inserting (\ref{eq:14}) in (\ref{eq:13}) the grand-canonical potential
$\Omega{}_{\mathrm{eff}}$ reduces up to first order in all fluctuations
to the grand-canonical free energy 
\begin{alignat}{1}
\mathcal{F}= & -\frac{v\mu^{2}}{2V_{\mathbf{0}}}\nonumber \\
 & +\frac{1}{2}\sum_{\mathbf{k}}^{\prime}\left[E_{\mathbf{k}}-\left(\frac{\hbar^{2}\mathbf{k}^{2}}{2m}+\mu\frac{V_{\mathbf{k}}}{V_{\mathbf{0}}}\right)\right]\nonumber \\
 & +\sum_{\mathbf{k}}^{\prime}\frac{1}{\beta}\mathrm{ln}\left(1-e^{-\beta E_{\mathbf{k}}}\right)\nonumber \\
 & -\sum_{\mathbf{k}}^{\prime}\frac{R_{\mathbf{k}}}{E_{\mathbf{k}}^{2}}\frac{\hbar^{2}\mathbf{k}^{2}}{2m}\frac{\mu}{V_{\mathbf{0}}}.\label{eq:15-1}
\end{alignat}
The particle number density follows then from equation (\ref{eq:15-1})
via the thermodynamic relation $n=-\frac{1}{v}\frac{\partial\mathcal{F}}{\partial\mu}$:
\begin{alignat}{1}
n= & \frac{\mu}{V_{\mathbf{0}}}-\frac{1}{2v}\sum_{\mathbf{k}}^{\prime}\left(\frac{\frac{\hbar^{2}\mathbf{k}^{2}}{2m}\frac{V_{\mathbf{k}}}{V_{\mathbf{0}}}}{E_{\mathbf{k}}}-\frac{V_{\mathbf{k}}}{V_{\mathbf{0}}}\right)\nonumber \\
 & -\frac{1}{v}\sum_{\mathbf{k}}^{\prime}\frac{\frac{\hbar^{2}\mathbf{k}^{2}}{2m}\frac{V_{\mathbf{k}}}{V_{\mathbf{0}}}}{\left(e^{\beta E_{\mathbf{k}}}-1\right)E_{\mathbf{k}}}\nonumber \\
 & +\frac{1}{v}\sum_{\mathbf{k}}^{\prime}\frac{R_{\mathbf{k}}}{E_{\mathbf{k}}^{4}}\frac{1}{V_{\mathbf{0}}}\left(\frac{\hbar^{2}\mathbf{k}^{2}}{2m}\right)^{3}.\label{eq:17}
\end{alignat}
Eliminating from (\ref{eq:14}) and (\ref{eq:17}) the chemical potential
$\mu$ allows to determine the condensate depletion up to first order
in the respective fluctuations:
\begin{alignat}{1}
n-n_{\mathbf{0}}= & n^{\prime}+n_{\mathrm{th}}+n_{\mathrm{R}}.\label{eq:17-1}
\end{alignat}
Thus the total particle density consists of four parts. Within the
Bogoliubov theory the main contribution is due to the condensate density
$n_{\mathbf{0}}$, whereas the occupation of the excited states above
the ground state consists of three distinct terms: the condensate
depletion due to interaction and quantum fluctuations is denoted by
\cite{yang}
\begin{alignat}{1}
n^{\prime}= & \frac{1}{2v}\sum_{\mathbf{k}}^{\prime}\left(\frac{\frac{\hbar^{2}\mathbf{k}^{2}}{2m}+nV_{\mathbf{k}}}{E_{\mathbf{k}}}-1\right).\label{eq:18}
\end{alignat}
Note that we neglected here any impact of the external random potential
upon the quantum fluctuations, an interesting case which was only
recently studied in detail in Refs.~\cite{muller,muller1}.

The condensate depletion due to the thermal fluctuations is given
by 
\begin{alignat}{1}
n_{\mathrm{th}}= & \frac{1}{v}\sum_{\mathbf{k}}^{\prime}\frac{\frac{\hbar^{2}\mathbf{k}^{2}}{2m}+nV_{\mathbf{k}}}{E_{\mathbf{k}}}\,\frac{1}{e^{\beta E_{\mathbf{k}}}-1}.\label{eq:19-1}
\end{alignat}
And finally, the condensate depletion due to the external random potential
results in
\begin{alignat}{1}
n_{\mathrm{R}}= & \frac{1}{v}\sum_{\mathbf{k}}^{\prime}\frac{nR_{\mathbf{k}}}{E_{\mathbf{k}}^{4}}\left(\frac{\hbar^{2}\mathbf{k}^{2}}{2m}\right)^{2}.\label{eq:20}
\end{alignat}
Note that the validity of our approximation depends on the condition
$n-n_{\mathbf{0}}\ll n$, i.e. the depletion is small, so that condensate
density $n_{\mathbf{0}}$ and total density $n$ are approximately
the same. With this dispersion (\ref{eq:16}) in all fluctuation terms
(\ref{eq:18})--(\ref{eq:20}) reduces to 
\begin{alignat}{1}
E_{\mathbf{k}}=\sqrt{\left(\frac{\hbar^{2}\mathbf{k}^{2}}{2m}\right)^{2}+nV_{\mathbf{k}}\frac{\hbar^{2}\mathbf{k}^{2}}{m}} & ,\label{eq:21}
\end{alignat}
which represents the celebrated Bogoliubov spectrum for the collective
excitations. Note that in the Popov approximation the particle density
$n$ in (\ref{eq:21}) is substituted by the condensate density $n_{\mathbf{0}}$,
so that this Bogoliubov dispersion acquires fluctuation corrections
\cite{jandersen}.

The grand-canonical ground-state energy $E-\mu N$ follows directly
from equation (\ref{eq:15-1}), by using the thermodynamic relation
$E-\mu N=-T^{2}\left(\frac{\partial}{\partial T}\frac{\mathcal{F}}{T}\right)_{v,\mu}$:
\begin{alignat}{1}
E-\mu N=-\frac{v\mu^{2}}{2V_{\mathbf{0}}}+E^{\prime}+E_{\mathrm{th}}+E_{\mathrm{R}} & .\label{eq:16-1}
\end{alignat}
The first term on the right-hand side of (\ref{eq:16-1}) represents
the ground-state mean-field energy for the homogeneous Bose-Einstein
condensate, whereas the other three terms denote energy contributions
which are due to the partial occupation of the excited states. The
energy shift due to interaction and quantum fluctuations is denoted
by
\begin{alignat}{1}
E^{\prime}=\frac{1}{2}\sum_{\mathbf{k}}^{\prime}\left[E_{\mathbf{k}}-\left(\frac{\hbar^{2}\mathbf{k}^{2}}{2m}+nV_{\mathbf{k}}\right)\right] & .\label{eq:22}
\end{alignat}
The energy shift due to the thermal fluctuations is given by
\begin{alignat}{1}
E_{\mathrm{th}}=\sum_{\mathbf{k}}^{\prime}E_{\mathbf{k}}\frac{1}{e^{\beta E_{\mathbf{k}}}-1} & .\label{eq:24}
\end{alignat}
And, finally, the energy shift due to the external random potential
reads
\begin{alignat}{1}
E_{\mathrm{R}}=-\sum_{\mathbf{k}}^{\prime}\frac{R_{\mathbf{k}}}{E_{\mathbf{k}}^{2}}\frac{\hbar^{2}\mathbf{k}^{2}}{2m}n & .\label{eq:23-1}
\end{alignat}
Within the next section these general results are further evaluated
and discussed for a dipolar Bose gas.

\begin{figure}[t]
\centering{}\includegraphics[scale=0.8]{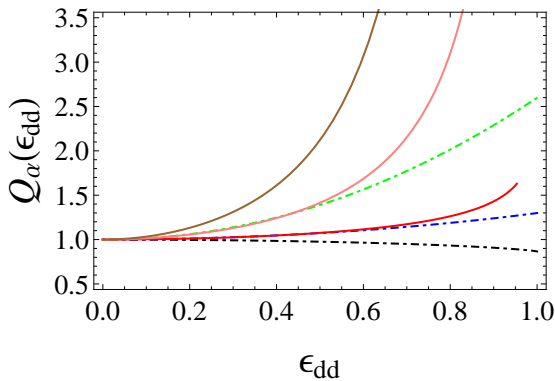}\protect\caption{\label{fig:The-dipolar-enhancement}(color online). Dipolar enhancement function $Q_{\alpha}(\epsilon_{\mathrm{dd}})$
versus relative dipolar interaction strength $\epsilon_{\mathrm{dd}}$
for different values of $\alpha$: -5/2 (brown, upper solid), -3/2
(pink, middle solid), -1/2 (red, lower solid), 1/2 (black, lower dotted-dashed),
3/2 (blue, middle dotted-dashed), 5/2 (green, upper dotted-dashed).}
\end{figure}

\section{Zero-Temperature results}

In this section we specialize our results to the zero-temperature
limit for a delta-correlated random potential $R(\mathbf{x}-\mathbf{x}^{\prime})=R_{0}\delta(\mathbf{x}-\mathbf{x}^{\prime})$,
where $R_{0}$ denotes the disorder strength. In the thermodynamic
limit, when both the particle number $N$ and the volume $v$ diverge,
i.e. $N\rightarrow\infty$ and $v\rightarrow\infty$, but their ratio
$n=\frac{N}{v}$ remains constant, the respective wave vector sums
converge towards integrals: $\frac{1}{v}\sum_{\mathbf{k}}^{\prime}\rightarrow\int\frac{d^{3}\mathbf{k}}{(2\pi)^{3}}$.
With this we obtain for the quantum depletion (\ref{eq:18}) the concrete
result \cite{key-15,lima} 
\begin{alignat}{1}
n^{\prime}=n_{\mathrm{B}}Q_{\frac{3}{2}}(\epsilon_{\mathrm{dd}}) & ,\label{eq:25}
\end{alignat}
with the original Bogoliubov expression \cite{bogoliubov}
\begin{alignat}{1}
n_{\mathrm{B}}=\frac{8}{3\sqrt{\pi}}\left(na\right)^{\frac{3}{2}} & ,\label{eq:29}
\end{alignat}
whereas the disorder depletion (\ref{eq:20}) specializes to \cite{pelster1,pelster3}
\begin{alignat}{1}
n_{\mathrm{R}}=n_{\mathrm{HM}}Q_{-\frac{1}{2}}(\epsilon_{\mathrm{dd}}) & .\label{eq:26}
\end{alignat}
Here $\epsilon_{\mathrm{dd}}=C_{\mathrm{dd}}/3g$ denotes the relative
dipolar interaction strength, and 
\begin{alignat}{1}
n_{\mathrm{HM}}=\frac{m^{2}R_{0}}{8\hbar^{4}\pi^{\frac{3}{2}}}\sqrt{\frac{n}{a}} & ,\label{eq:31}
\end{alignat}
represents the Huang-Meng result for the isotropic contact interaction
\cite{key-11}. In addition, the functions

\begin{alignat}{1}
Q_{\alpha}(\epsilon_{\mathrm{dd}})=\int_{0}^{1}\,\left[1+\epsilon_{\mathrm{dd}}(3x^{2}-1)\right]^{\alpha}dx & ,\label{eq:23}
\end{alignat}
which describe the dipolar effect, can be expressed analytically \cite{gradshtien}
\begin{alignat}{1}
Q_{\alpha}(\epsilon_{\mathrm{dd}})= & (1-\epsilon_{\mathrm{dd}})^{\alpha}\,_{2}F_{1}\left(-\alpha,\frac{1}{2};\frac{3}{2};\frac{-3\epsilon_{\mathrm{dd}}}{1-\epsilon_{\mathrm{dd}}}\right)
\end{alignat}
in terms of the hypergeometric function $_{2}F_{1}$. The dipolar
function $Q_{\alpha}(\epsilon_{\mathrm{dd}})$ is depicted in Fig.~\ref{fig:The-dipolar-enhancement},
and increases for increasing magnitude of $\alpha$ for both positive
and negative sign. Note that, in order to avoid any instability, we
restricted $\epsilon_{\mathrm{dd}}$ in Fig.~\ref{fig:The-dipolar-enhancement}
to the maximum value one, so that the radicand in the Bogoliubov spectrum
(\ref{eq:21}) remains positive when $\mathbf{k}\rightarrow0$ \cite{key-15,lima}.

We conclude that Eqs.~(\ref{eq:25}), (\ref{eq:26}) reproduce the
contact interaction results Eqs.~(\ref{eq:29}), (\ref{eq:31}) for
$\epsilon_{\mathrm{dd}}=0$ due to the property $Q_{\alpha}(0)=1$.
Note that increasing the repulsive contact interaction has opposite
effects on the zero-temperature condensate depletion. On the one hand
more and more bosons are then scattered from the ground state to some
excited states so that the quantum depletion (\ref{eq:29}) increases.
On the other hand the localization of bosons in the respective minima
of the random potential is hampered, so that disorder depletion (\ref{eq:31})
decreases. In contrast to that the long-range dipolar interaction
turns out to enhance both the quantum deletion (\ref{eq:25}) and
the disorder depletion (\ref{eq:26}), see Fig.~\ref{fig:The-dipolar-enhancement}.
As the dipolar interaction potential (\ref{eq:3}) has two repulsive
and only one attractive direction in real space, it yields a net expulsion
of bosons from the ground state which is described by $Q_{\frac{3}{2}}(\epsilon_{\mathrm{dd}})$
in (\ref{eq:25}). In contrast to that the dipolar interaction supports
the localization of bosons in the random environment \cite{pelster1,pelster3}
according to $Q_{-\frac{1}{2}}(\epsilon_{\mathrm{dd}})$ in (\ref{eq:26}).
This is due to the fact that the BEC droplets in the respective minima
of the random potential can minimize their energy by deformation along
the polarization axis \cite{pfau-3}. The dipolar enhancement in both
the quantum depletion $n^{\prime}$ and the external random potential
depletion $n_{\mathrm{R}}$ are approximately the same with respect
to $\epsilon_{\mathrm{dd}}$ up to $\epsilon_{\mathrm{dd}}=0.6$,
but then the dipolar enhancement in $n_{\mathrm{R}}$ starts to increase
faster than in $n^{\prime}$, see Fig.~\ref{fig:The-dipolar-enhancement}.

Furthermore, we read off from (\ref{eq:22}) that the energy shift
due to the interaction and the quantum fluctuations are ultraviolet
divergent. Thus, following Refs.~\cite{lima,fetter}, we can evaluate
(\ref{eq:22}) by introducing an ultraviolet cutoff, where the interaction
has to be renormalized by inserting the term $\frac{m(nV_{\mathbf{k}})^{2}}{\hbar^{2}\mathbf{k}^{2}}$,
so that the divergent part is removed 
\begin{alignat}{1}
\frac{E^{\prime}}{N}=\frac{1}{2n}\int\frac{d^{3}\mathbf{k}}{(2\pi)^{3}}\,\left[E_{\mathbf{k}}-\left(\frac{\hbar^{2}\mathbf{k}^{2}}{2m}+nV_{\mathbf{k}}\right)+\frac{m(nV_{\mathbf{k}})^{2}}{\hbar^{2}\mathbf{k}^{2}}\right] & ,
\end{alignat}
 evaluations in the thermodynamic limit yielding 
\begin{alignat}{1}
\frac{E^{\prime}}{N}=\frac{2\pi a\hbar^{2}n}{m}\frac{128}{15}\left(\frac{na^{3}}{\pi}\right)^{\frac{1}{2}}Q_{\frac{5}{2}}(\epsilon_{\mathrm{dd}}) & .\label{eq:30-1}
\end{alignat}
Another method for (\ref{eq:22}) relies on Schwinger trick \cite{kleinert}
which leads again to (\ref{eq:30-1}). The energy shift due to the
external random potential (\ref{eq:23-1}) is also ultraviolet divergent:
\begin{alignat}{1}
\frac{E_{\mathrm{R}}}{N}=-\int\frac{d^{3}\mathbf{k}}{(2\pi)^{3}}\,\frac{R_{0}}{\frac{\hbar^{2}\mathbf{k}^{2}}{2m}+2nV_{\mathbf{k}}} & .
\end{alignat}
Thus, it is calculated either with an ultraviolet cutoff by subtracting
the divergent term $\frac{2mR_{0}}{v\hbar^{2}\mathbf{k}^{2}}$, or
with the help of the Schwinger trick, leading to
\begin{alignat}{1}
\frac{E_{\mathrm{R}}}{N}= & \frac{2mR_{0}}{\hbar^{2}}\sqrt{\frac{na}{\pi}}Q_{\frac{1}{2}}(\epsilon_{\mathrm{dd}}).\label{eq:32}
\end{alignat}
Note that energy shift $E^{\prime}$ due to interaction and quantum
fluctuations in equation (\ref{eq:30-1}) and the external random
potential energy shift $E_{\mathrm{R}}$ in equation (\ref{eq:32})
reproduce the well-known contact interaction results when $\epsilon_{\mathrm{dd}}=0$,
and that $E^{\prime}$ increases quicker than $E_{\mathrm{R}}$ with
respect to $\epsilon_{\mathrm{dd}}$. Finally note that both energy
shifts (\ref{eq:30-1}) and (\ref{eq:32}) could be used to study
the collective properties of a trapped dipolar Bose gas within the
local density approximation \cite{huang,huang1}.

\section{Superfluidity}

Now we extend the Bogoliubov theory in order to study the transport
phenomenon of superfluidity within the framework of linear response
theory. With this we calculate the superfluid depletions due to both
the external random potential and the thermal fluctuations.

\subsection{Bogoliubov theory revisited}

Following Refs.~\cite{key-12,ueda} we calculate the superfluid component
$n_{s}$ of the Bose gas, which moves with the superfluid velocity
$\mathbf{v}_{\mathrm{s}}$, via a linear response approach. To this
end we perform a Galilean boost for our system with a boost velocity
$\mathbf{u}$ so that the normal fluid component $n_{n}$ is dragged
along this boost. Inserting the Galilean transformations $\mathbf{x}^{\prime}=\mathbf{x}+\mathbf{u}t$,
$t=t^{\prime}$, and rewriting the field operator in Heisenberg picture
according to $\hat{\psi}^{\prime}(\mathbf{x}^{\prime},t^{\prime})=\hat{\psi}(\mathbf{x},t)e^{\frac{\imath}{\hbar}m\mathbf{v}_{\mathrm{s}}\mathbf{x}}$,
the Hamiltonian in momentum space reads instead of (\ref{eq:4})
\begin{alignat}{1}
\mathcal{\hat{K}}= & \frac{1}{2}\sum_{\mathbf{k}}\left[\frac{\hbar^{2}\mathbf{k}^{2}}{2m}-\mu_{\mathrm{eff}}\right](\hat{a}_{\mathbf{k}}^{\dagger}\hat{a}_{\mathbf{k}}+\hat{a}_{-\mathbf{k}}^{\dagger}\hat{a}_{-\mathbf{k}})\nonumber \\
 & +\frac{1}{2}\sum_{\mathbf{k}}\hbar\mathbf{k}(\mathbf{u}-\mathbf{v}_{\mathrm{s}})(\hat{a}_{\mathbf{k}}^{\dagger}\hat{a}_{\mathbf{k}}-\hat{a}_{-\mathbf{k}}^{\dagger}\hat{a}_{-\mathbf{k}})\nonumber \\
 & +\frac{1}{2v}\sum_{\mathbf{p},\mathbf{k}}U_{\mathbf{p}-\mathbf{k}}(\hat{a}_{\mathbf{p}}^{\dagger}\hat{a}_{\mathbf{k}}+\hat{a}_{-\mathbf{k}}^{\dagger}\hat{a}_{-\mathbf{p}})\nonumber \\
 & +\frac{1}{2v}\sum_{\mathbf{p},\mathbf{k},\mathbf{q}}V_{\mathbf{q}}\:\hat{a}_{\mathbf{k}+\mathbf{q}}^{\dagger}\hat{a}_{\mathbf{p}-\mathbf{q}}^{\dagger}\hat{a}_{\mathbf{p}}\hat{a}_{\mathbf{k}},\label{eq:27}
\end{alignat}
with the effective chemical potential $\mu_{\mathrm{eff}}=\mu-\frac{1}{2}m\mathbf{v}_{\mathrm{s}}^{2}+m\mathbf{u}\mathbf{v}_{\mathrm{s}}$.
Applying the Bogoliubov theory along similar lines as in Section II
and after performing the disorder ensemble average of the grand-canonical
Hamiltonian we obtain instead of (\ref{eq:12})
\begin{alignat}{1}
\left\langle \mathcal{\hat{K}}^{\prime}\right\rangle = & v\left(-\mu_{\mathrm{eff}}n_{\mathbf{0}}+\frac{1}{2}V_{\mathbf{0}}n_{\mathbf{0}}^{2}\right)\nonumber \\
 & +\frac{1}{2}\sum_{\mathbf{k}}^{\prime}\left\{ E_{\mathbf{k}}-\left[\frac{\hbar^{2}\mathbf{k}^{2}}{2m}-\mu_{\mathrm{eff}}+n_{\mathbf{0}}(V_{\mathbf{0}}+V_{\mathbf{k}})\right]\right\} \nonumber \\
 & +\sum_{\mathbf{k}}^{\prime}\left[E_{\mathbf{k}}+\hbar\mathbf{k}(\mathbf{u}-\mathbf{v}_{\mathrm{s}})\right]\hat{\alpha}_{\mathbf{k}}^{\dagger}\hat{\alpha}_{\mathbf{k}}\nonumber \\
 & -\sum_{\mathbf{k}}^{\prime}\frac{n_{\mathbf{0}}R_{\mathbf{k}}(\frac{\hbar^{2}\mathbf{k}^{2}}{2m}-\mu_{\mathrm{eff}}+n_{\mathbf{0}}V_{0})}{\left\{ E_{\mathbf{k}}^{2}-\left[\hbar\mathbf{k}(\mathbf{u}-\mathbf{v}_{\mathrm{s}})\right]^{2}\right\} },
\end{alignat}
where $E_{\mathbf{k}}$ stands for the quasi-particle dispersion (\ref{eq:10-1})
with $\mu$ substituted by $\mu_{\mathrm{eff}}$. Now it is straightforward
to find the corresponding grand-canonical effective potential, where
we get instead of (\ref{eq:13}):
\begin{alignat}{1}
\Omega{}_{\mathrm{eff}}= & v\left(-\mu_{\mathrm{eff}}n_{\mathbf{0}}+\frac{1}{2}V_{\mathbf{0}}n_{\mathbf{0}}^{2}\right)\nonumber \\
 & +\frac{1}{2}\sum_{\mathbf{k}}^{\prime}\left\{ E_{\mathbf{k}}-\left[\frac{\hbar^{2}\mathbf{k}^{2}}{2m}-\mu_{\mathrm{eff}}+n_{\mathbf{0}}(V_{\mathbf{0}}+V_{\mathbf{k}})\right]\right\} \nonumber \\
 & +\sum_{\mathbf{k}}^{\prime}\frac{1}{\beta}\mathrm{ln}\left\{ 1-e^{-\beta\left[E_{\mathbf{k}}+\hbar\mathbf{k}(\mathbf{u}-\mathbf{v}_{\mathrm{s}})\right]}\right\} \nonumber \\
 & -\sum_{\mathbf{k}}^{\prime}\frac{n_{\mathbf{0}}R_{\mathbf{k}}\left(\frac{\hbar^{2}\mathbf{k}^{2}}{2m}-\mu_{\mathrm{eff}}+n_{\mathbf{0}}V_{\mathbf{0}}\right)}{\left\{ E_{\mathbf{k}}^{2}-\left[\hbar\mathbf{k}(\mathbf{u}-\mathbf{v}_{\mathrm{s}})\right]^{2}\right\} }.\label{eq:19}
\end{alignat}
Within a linear response approach we are only interested in small
velocities $\mathbf{v}_{\mathrm{s}}$ and $\mathbf{u}$, so we expand
Eq.~(\ref{eq:19}) up to second order in $\hbar\mathbf{k\left(u-\mathbf{v}_{\mathrm{s}}\right)}$,
and we get
\begin{alignat}{1}
\Omega{}_{\mathrm{eff}}= & v\left(-\mu_{\mathrm{eff}}n_{\mathbf{0}}+\frac{1}{2}V_{\mathbf{0}}n_{\mathbf{0}}^{2}\right)\nonumber \\
 & +\frac{1}{2}\sum_{\mathbf{k}}^{\prime}\left\{ E_{\mathbf{k}}-\left[\frac{\hbar^{2}\mathbf{k}^{2}}{2m}-\mu_{\mathrm{eff}}+n_{\mathbf{0}}(V_{\mathbf{0}}+V_{\mathbf{k}})\right]\right\} \nonumber \\
 & +\sum_{\mathbf{k}}^{\prime}\frac{1}{\beta}\mathrm{ln}\left(1-e^{-\beta E_{\mathbf{k}}}\right)-\sum_{\mathbf{k}}^{\prime}\frac{\beta e^{\beta E_{\mathbf{k}}}\left[\hbar\mathbf{k}(\mathbf{u}-\mathbf{v}_{\mathrm{s}})\right]^{2}}{2\left(e^{\beta E_{\mathbf{k}}}-1\right)^{2}}\nonumber \\
 & -\sum_{\mathbf{k}}^{\prime}\frac{n_{\mathbf{0}}R_{\mathbf{k}}}{E_{\mathbf{k}}^{2}}\left[\frac{\hbar^{2}\mathbf{k}^{2}}{2m}-\mu_{\mathrm{eff}}+n_{\mathbf{0}}V_{\mathbf{0}}\right]\label{eq:30}\\
 & -\sum_{\mathbf{k}}^{\prime}\frac{n_{\mathbf{0}}R_{\mathbf{k}}}{E_{\mathbf{k}}^{4}}\left[\frac{\hbar^{2}\mathbf{k}^{2}}{2m}-\mu_{\mathrm{eff}}+n_{\mathbf{0}}V_{\mathbf{0}}\right]\left[\hbar\mathbf{k}(\mathbf{u}-\mathbf{v}_{\mathrm{s}})\right]^{2}.\nonumber 
\end{alignat}
Note that the terms linear in $\hbar\mathbf{k\left(u-\mathbf{v}_{\mathrm{s}}\right)}$
vanish due to symmetry reasons. Extremizing the grand-canonical effective
potential (\ref{eq:30}) with respect the the condensate density $n_{\mathbf{0}}$
yields in zeroth order the mean-field result $\mu_{\mathrm{eff}}=n_{\mathbf{0}}V_{\mathbf{0}}$.
With this we are ready to determine the momentum of the system $\mathrm{\mathbf{P}}=\left(-\frac{\partial\Omega{}_{\mathrm{eff}}}{\partial\mathbf{u}}\right)_{v,T,\mu}$
in the small-velocity limit, which turns out to be of the following
form

\begin{alignat}{1}
\mathrm{\mathbf{P}}=mv\left(n \mathbf{v}_{s}+n_{\mathrm{n}}\mathbf{v}_{\mathrm{n}}\right) & ,\label{eq:38}
\end{alignat}
where the total density $n=n_{\mathrm{s}}+n_{\mathrm{n}}$ decomposes
into the superfluid and normal component $n_{\mathrm{s}}$ and $n_{\mathrm{n}}$,
respectively, and $\mathbf{v}_{\mathrm{n}}=\mathbf{u}-\mathbf{v}_{\mathrm{s}}$
denotes the normal fluid velocity. The normal fluid density turns
out to decompose according to $n_{\mathrm{n}}=n_{\mathrm{R}}+n_{\mathrm{th}}$,
where the contribution due to the external random potential, which
only exists in the zero temperature limit, reads \cite{pelster1}
\begin{alignat}{1}
n_{\mathrm{R}ij}=\frac{1}{v}\sum_{\mathbf{k}}^{\prime}\frac{2nR_{\mathbf{k}}\hbar^{2}k_{i}k_{j}}{m(\frac{\hbar^{2}\mathbf{k}^{2}}{2m})\left(\frac{\hbar^{2}\mathbf{k}^{2}}{2m}+2nV_{\mathbf{k}}\right)^{2}} & ,\label{eq:34}
\end{alignat}
while the normal fluid density due to the thermal fluctuations is
given by
\begin{alignat}{1}
n_{\mathrm{th}ij}=\frac{1}{v}\sum_{\mathbf{k}}^{\prime}\frac{\beta}{m}\hbar^{2}k_{i}k_{j}\frac{e^{\beta E_{\mathbf{k}}}}{\left(e^{\beta E_{\mathbf{k}}}-1\right)^{2}} & ,\label{eq:40}
\end{alignat}
where $E_{\mathbf{k}}$ stands for the Bogoliubov energy spectrum
(\ref{eq:21}). Note that there is no superfluid depletion in (\ref{eq:38})
which is due to quantum fluctuations, in contrast to the condensate
depletion which has a quantum fluctuations component determined by
Eq.~(\ref{eq:18}).

\subsection{Zero-temperature superfluid  depletion}

\begin{figure}[t]
\begin{centering}
\includegraphics[scale=0.8]{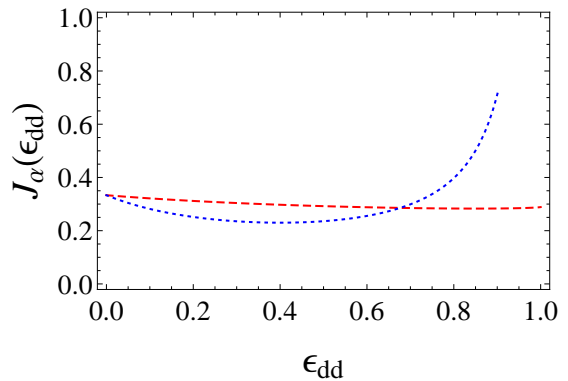}
\par\end{centering}
\protect\caption{\label{fig:The-functionJ}(color online). Function $J_{\alpha}(\epsilon_{\mathrm{dd}})$
versus relative dipolar interaction strength $\epsilon_{\mathrm{dd}}$
for different values of $\alpha$: -1/2 (red, dashed), -5/2 (blue,
dotted).}
\end{figure}

Again we specialize at first to the zero-temperature limit and to
a delta-correlated random potential and evaluate the corresponding
superfluid depletions in the thermodynamic limit. Depending on the
boost direction, we have two different superfluid depletions in the
directions parallel or perpendicular to the dipole polarization. In
the first case the superfluid depletion reads 
\begin{alignat}{1}
n_{\mathrm{R}\parallel}=4n_{\mathrm{HM}}\: J_{-\frac{1}{2}}(\epsilon_{\mathrm{dd}}) & ,\label{eq:43}
\end{alignat}
whereas in the second case the superfluid depletion turns out to be
\begin{alignat}{1}
n_{\mathrm{R}\perp}=2n_{\mathrm{HM}}\,\left[Q_{-\frac{1}{2}}(\epsilon_{\mathrm{dd}})-J_{-\frac{1}{2}}(\epsilon_{\mathrm{dd}})\right] & .\label{eq:44}
\end{alignat}
Here we have introduced the new function
\begin{alignat}{1}
J_{\alpha}(\epsilon_{\mathrm{dd}})=\int_{0}^{1}\, x^{2}\,\left[1+\epsilon_{\mathrm{dd}}(3x^{2}-1)\right]{}^{\alpha}dx & ,\label{eq:42}
\end{alignat}
which can also be expressed analytically with the help of the integral
table \cite{gradshtien} according to
\begin{alignat}{1}
J_{\alpha}(\epsilon_{\mathrm{dd}})=\frac{1}{3}(1-\epsilon_{\mathrm{dd}})^{\alpha}\,{}_{2}F_{1}\left(-\alpha,\frac{3}{2};\frac{5}{2};\frac{-3\epsilon_{\mathrm{dd}}}{1-\epsilon_{\mathrm{dd}}}\right) & .
\end{alignat}
Note that the function $J_{\alpha}(\epsilon_{\mathrm{dd}})$ shown
in Fig.~\ref{fig:The-functionJ} has the property $J_{\alpha}(0)=1/3$
and decreases slowly with increasing $\epsilon_{\mathrm{dd}}$ for
$\alpha=-\frac{1}{2}$, while it behaves non-monotonically with increasing
$\epsilon_{\mathrm{dd}}$ for $\alpha=-\frac{5}{2}$.

\begin{figure}[t]
\begin{centering}
\includegraphics[scale=0.45]{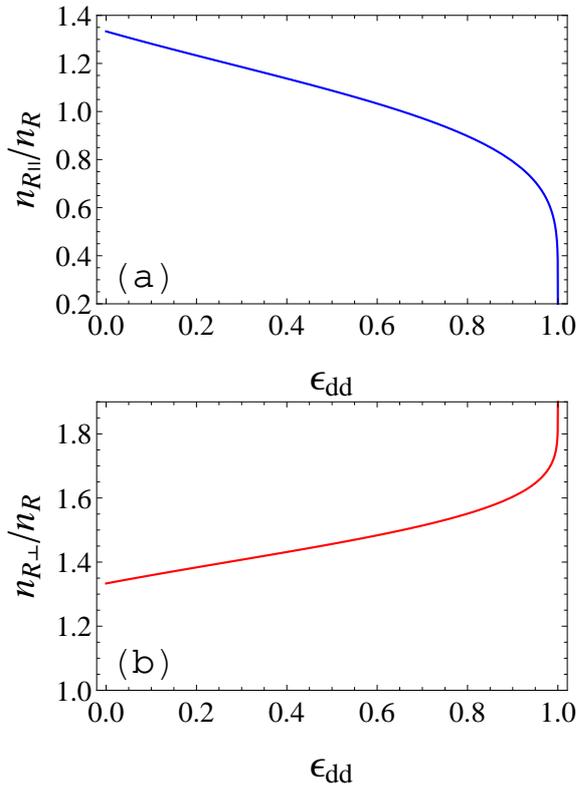}
\par\end{centering}
\protect\caption{\label{superfluid1ratio}(color online). Ratios of superfluid depletions $n_{\mathrm{R}\parallel}$
and $n_{\mathrm{R}\perp}$ and condensate depletion $n_{\mathrm{R}}$
versus relative dipolar interaction strength $\epsilon_{\mathrm{dd}}$.}
\end{figure}

We observe that, due to $Q_{\alpha}(0)=1$ and $J_{\alpha}(0)=1/3$,
both depletions $n_{\mathrm{R}\parallel}$ and $n_{\mathrm{R}\perp}$
coincide for vanishing dipolar interaction and equate $\frac{4}{3}n_{\mathrm{HM}}$,
which reproduces the Huang-Meng result for the isotropic contact interaction
\cite{key-11}. Due to the localization of bosons in the respective
minima of the random potential the superfluid is hampered, so that
the superfluid depletion is larger than the condensate depletion.
Switching on the dipolar interaction has a significant effect on both
depletions $n_{\mathrm{R}\parallel}$ and $n_{\mathrm{R}\perp}$.
Their ratio with the condensate depletion $n_{\mathrm{R}}$ turns
out to be monotonically decreasing/increasing with respect to the
relative dipolar interaction strength $\epsilon_{\mathrm{dd}}$ according
to Fig.~\ref{superfluid1ratio}. This is explained qualitatively
by considering the dipolar interaction in Fourier space (\ref{eq:92})
as an effective contact interaction \cite{pelster1,pelster3}. Therefore,
for sufficiently large values of $\epsilon_{\mathrm{dd}}$, it turns
out that $n_{\mathrm{R}\parallel}$ becomes even smaller than the
condensate depletion. This surprising finding suggests that the localization
of the bosons in the minima of the random potential only occurs for
a finite period of time \cite{graham}. As a consequence the ratio
$n_{\mathrm{R}\parallel}/n_{\mathrm{R}\perp}$ decreases monotonically
for increasing values of the relative dipolar interaction strength
$\epsilon_{\mathrm{dd}}$.

\section{Finite-temperature effects}

In the previous sections we restricted the application of the Bogoliubov
theory of dirty bosons to the zero-temperature limit. Nevertheless
any experiment is performed at finite-temperature, where thermal fluctuations
play an important role besides quantum and disorder fluctuations.
Therefore, we calculate in this section for a dipolar BEC the contributions
to both the condensate and the normal fluid component which are due
to thermal excitations.

\begin{figure}[t]
\begin{centering}
\includegraphics[scale=0.8]{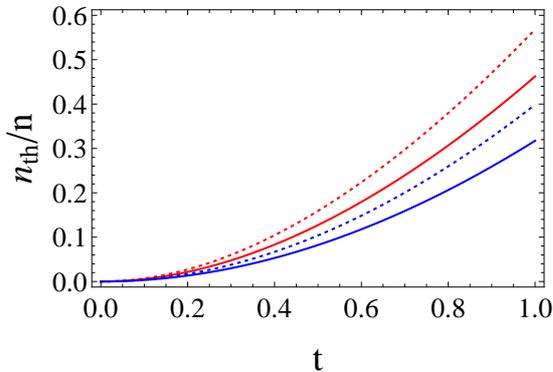}
\par\end{centering}
\protect\caption{\label{fig:The-thermal-fractional}(color online). Thermal fractional depletion $n_{\mathrm{th}}/n$
versus relative temperature $t=T/T_{c}^{0}$ for different values of relative
dipolar interaction strength $\epsilon_{\mathrm{dd}}=0$ (solid),
$\epsilon_{\mathrm{dd}}=0.8$ (dotted), and gas parameter $\gamma=0.01$
(red, upper two curves), $\gamma=0.20$ (blue).}
\end{figure}

\subsection{Condensate depletion}

At zero temperature the condensate depletion consists of two contributions,
one due to the quantum fluctuation (\ref{eq:18}) and another one
due to the disorder potential (\ref{eq:20}), which are evaluated
in (\ref{eq:25}), (\ref{eq:26}) for a dipolar BEC, respectively.
But for increasing temperature also the third contribution (\ref{eq:19-1})
of the depletion (\ref{eq:17-1}) will play an important role and
should be taken into account. This additional term represents the
condensate depletion due to the thermal excitations, which reads in
the thermodynamics limit

\begin{alignat}{1}
n_{\mathrm{th}}=\int\frac{d^{3}\mathbf{k}}{(2\pi)^{3}}\,\frac{\frac{\hbar^{2}\mathbf{k}^{2}}{2m}+nV_{\mathbf{k}}}{E_{\mathbf{k}}}\,\frac{1}{e^{\beta E_{\mathbf{k}}}-1} & ,\label{eq:48}
\end{alignat}
where $E_{\mathbf{k}}$ denotes the Bogoliubov spectrum (\ref{eq:21}).
The integral (\ref{eq:48}) can be recast in dimensionless form
\begin{alignat}{1}
\frac{n_{\mathrm{th}}}{n}=\frac{\gamma^{-\frac{1}{6}}t^{2}}{2\pi^{\frac{1}{2}}\left(\zeta(\frac{3}{2})\right)^{\frac{4}{3}}}I(\gamma,\epsilon_{\mathrm{dd}},t) & ,\label{eq:52}
\end{alignat}
where we have introduced the gas parameter $\gamma=na^{3}$, and the
relative temperature $t=T/T_{c}^{0}$, with $T_{c}^{0}=\frac{2\pi\hbar^{2}n^{\frac{2}{3}}}{\left(\zeta(\frac{3}{2})\right)^{\frac{2}{3}}mk_{\mathrm{B}}}$
being the critical temperature for the non-interacting Bose gas, and $\zeta(x)$ denotes the Riemann zeta function with the value $\zeta(\frac{3}{2})=2.61238..$. The
remaining integral $I(\gamma,\epsilon_{\mathrm{dd}},t)$ reads
\begin{alignat}{1}
I(\gamma,\epsilon_{\mathrm{dd}},t)=\int_{0}^{\infty}dx\int_{0}^{\pi}d\theta\frac{x\,\sin\theta\left(1+\frac{\alpha x^{2}}{8\varTheta^{2}}\right)}{\sqrt{\varTheta+\frac{\alpha x^{2}}{16\varTheta}}\left(e^{\sqrt{x^{2}+\frac{\alpha x^{4}}{16\varTheta^{2}}}}-1\right)} & ,\label{eq:45}
\end{alignat}
with the abbreviations $\alpha=\left[\frac{t}{\gamma^{\frac{1}{3}}\left(\zeta(\frac{3}{2})\right)^{\frac{2}{3}}}\right]^{2}$
and $\varTheta=1+\epsilon_{\mathrm{dd}}\left(3\cos^{2}\theta-1\right)$.
Fig.~\ref{fig:The-thermal-fractional} shows how the condensate thermal
fractional depletion (\ref{eq:52}) increases with the relative temperature
$t$ for different values of the relative dipolar interaction strength
$\epsilon_{\mathrm{dd}}$ and the gas parameter $\gamma$. For increasing
contact interaction, it turns out that $n_{\mathrm{th}}$ decreases
as collisions between the thermal particles yield partially a scattering
into the macroscopic ground state. In contrast to that we observe
that the dipolar interaction has the opposite effect: increasing $\epsilon_{\mathrm{dd}}$
enhances the thermal depletion $n_{\mathrm{th}}$. This can be understood
from considering the consequences of the dipolar interaction as an
effective contact interaction \cite{pelster1,pelster3}, yielding
two attractive and one repulsive direction in Fourier space according
to (\ref{eq:92}).

In order to investigate this qualitative finding more quantitatively
in the vicinity of zero temperature but far below the critical temperature,
we approximate the condensate depletion (\ref{eq:52}) analytically
like
\begin{alignat}{1}
\frac{n_{\mathrm{th}}}{n}= & \frac{\pi^{\frac{3}{2}}\gamma^{-\frac{1}{6}}t^{2}}{6\left(\zeta(\frac{3}{2})\right)^{\frac{4}{3}}}Q_{-\frac{1}{2}}(\epsilon_{\mathrm{dd}})\nonumber \\
 & -\frac{\pi^{\frac{7}{2}}\gamma^{-\frac{5}{6}}t^{4}}{480\left(\zeta(\frac{3}{2})\right)^{\frac{8}{3}}}Q_{-\frac{5}{2}}(\epsilon_{\mathrm{dd}})+...\quad.\label{eq:46}
\end{alignat}
This reproduces the isotropic contact interaction case for vanishing
dipolar interaction due to $Q_{\alpha}(0)=1$ \cite{landau-1}. Note
that (\ref{eq:46}) turns out to approximate (\ref{eq:45}) quite
well irrespective of $\gamma$ and $\epsilon_{\mathrm{dd}}$ in the
temperature range $0\leq t\leq0.3$. Here also each term in (\ref{eq:46})
decreases for increasing $\gamma$, while it is enhanced for increasing
$\epsilon_{\mathrm{dd}}$.

Now we investigate how the validity range of the Bogoliubov theory
depends on the gas parameter $\gamma$ and the relative temperature
$t$ for different values of the relative dipolar interaction strength
$\epsilon_{\mathrm{dd}}$ and the disorder strength $R_{0}$. To this
end we restrict the condensate depletion $\Delta n=n-n_{\mathbf{0}}$
in equation (\ref{eq:17-1}) to be limited maximally by half of the
particle number density $n$. Inserting the critical temperature for
the non-interacting Bose gas $T_{c}^{0}$ and the gas parameter $\gamma$
in the condensate depletion due to quantum fluctuation (\ref{eq:25})
and random potential (\ref{eq:26}), the fractional depletion reads

\begin{figure}[t]
\begin{centering}
\includegraphics[scale=0.8]{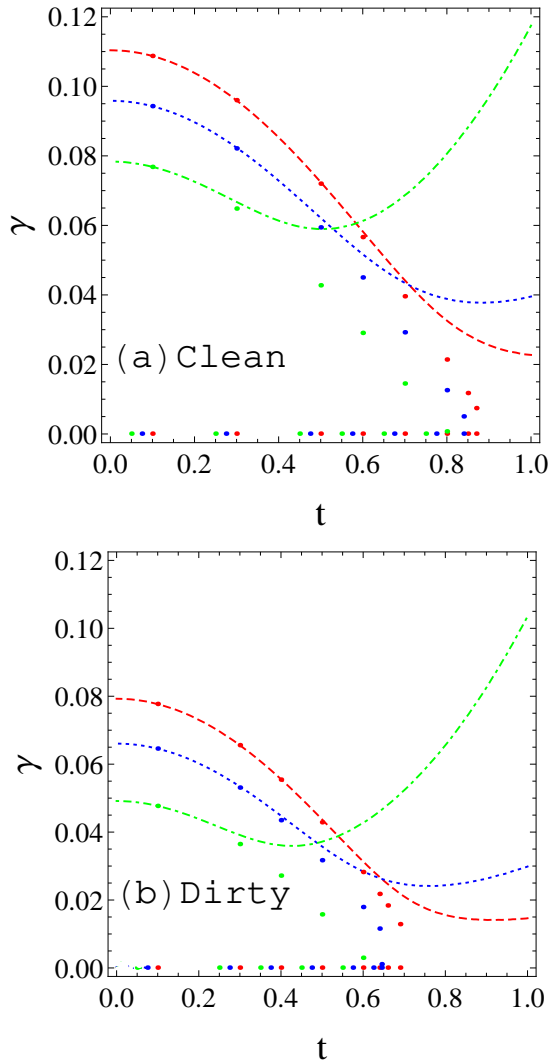}
\par\end{centering}
\protect\caption{\label{fig:The-gas-parameter}(color online). Validity range of Bogoliubov theory in the $t-\gamma$ plane for (a) clean case, i.e. $R_{0}=0$, and (b) dirty case with $R_{0}=\frac{2\hbar^{4}\pi^{\frac{3}{2}}n^{\frac{1}{3}}}{5m^{2}}$ for
different values of relative dipolar interaction strength $\epsilon_{\mathrm{dd}}=0$
(red, dashed), $\epsilon_{\mathrm{dd}}=0.5$ (blue, dotted), $\epsilon_{\mathrm{dd}}=0.8$
(green, dotted-dashed). Condensate depletion (\ref{eq:47}) equals $n/2$ for curves with
$n_{\mathrm{th}}/n$ from (\ref{eq:46}). Dots represent the
case upon inserting $n_{\mathrm{th}}/n$ from (\ref{eq:52}), (\ref{eq:45}).}
\end{figure}

\begin{alignat}{1}
\frac{\Delta n}{n}= & \frac{8}{3\sqrt{\pi}}\gamma^{\frac{1}{2}}Q_{\frac{3}{2}}(\epsilon_{\mathrm{dd}})+\frac{n_{\mathrm{th}}}{n}\nonumber \\
 & +\frac{m^{2}R_{0}}{8\hbar^{4}\pi^{\frac{3}{2}}n^{\frac{1}{3}}}\gamma^{-\frac{1}{6}}Q_{-\frac{1}{2}}(\epsilon_{\mathrm{dd}}).\label{eq:47}
\end{alignat}
Here the thermal fractional condensate depletion (\ref{eq:52}) in
(\ref{eq:47}) is determined numerically from (\ref{eq:45}) and from
(\ref{eq:46}) analytically in the vicinity of zero temperature for
comparison. Fig.~\ref{fig:The-gas-parameter} depicts the resulting
validity range of the Bogoliubov theory in the clean case i.e. the
disorder strength $R_{0}=0$, and in the dirty case with $R_{0}=\frac{2\hbar^{4}\pi^{\frac{3}{2}}n^{\frac{1}{3}}}{5m^{2}}$.
It is represented by the area below the curves when the condensate depletion
is equal to $n/2$. Outside this area the Bogoliubov approximation
is not valid. Note that the validity range (\ref{eq:52}), (\ref{eq:47})
is well approximated by (\ref{eq:46}) near zero temperature. From
(\ref{eq:52}), (\ref{eq:47}) we read off that the Bogoliubov theory
is not applicable near the critical temperature. Furthermore the validity
range decreases for increasing values of the relative dipolar interaction
strength $\epsilon_{\mathrm{dd}}$, and it also smaller in the dirty
case compared to the clean one.

\begin{figure}[t]
\centering{}\includegraphics[scale=0.8]{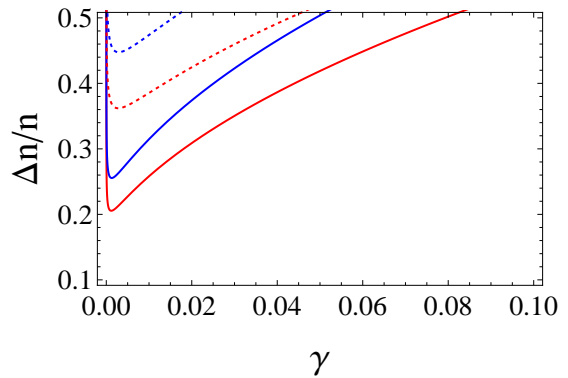}\protect\caption{\label{fig:Fractional-depletion-}(color online). Fractional depletion $\Delta n/n$
versus gas parameter $\gamma=na^{3}$ for different values of relative dipolar
interaction strength $\epsilon_{\mathrm{dd}}=0$ (red), $\epsilon_{\mathrm{dd}}=0.8$
(blue) and relative temperature $t=0$ (solid), $t=0.5$ (dotted)
with the disorder strength $R_{0}=\frac{2\hbar^{4}\pi^{\frac{3}{2}}n^{\frac{1}{3}}}{5m^{2}}$.}
\end{figure}

Finally, we plot in Fig.~\ref{fig:Fractional-depletion-} the total
fractional condensate depletion $\frac{\Delta n}{n}$ versus the gas
parameter $\gamma$ with the thermal fractional condensate depletion
$\frac{n_{\mathrm{th}}}{n}$ in (\ref{eq:47}) inserted from (\ref{eq:52})
for the disorder strength $R_{0}=\frac{2\hbar^{4}\pi^{\frac{3}{2}}n^{\frac{1}{3}}}{5m^{2}}$.
Note that the sharp increase for small gas parameter is unphysical
due to the fact that the Bogoliubov theory is no longer valid, see
Fig.~\ref{fig:The-gas-parameter}. We conclude that increasing the
parameters $t$ and $\epsilon_{\mathrm{dd}}$ yields an increasing
fractional depletion.

\begin{figure}[t]
\begin{centering}
\includegraphics[scale=0.45]{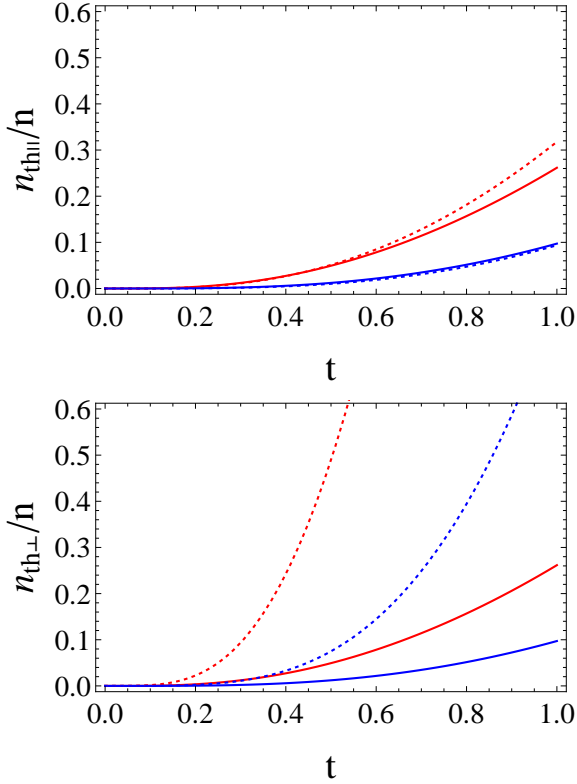}
\par\end{centering}
\protect\caption{\label{fig:The-superfluid-thermal}(color online). Superfluid thermal fractional depletions
$n_{\mathrm{th\Vert}}/n$ and $n_{\mathrm{th\bot}}/n$ versus relative
temperature $t=T/T_{c}^{0}$ for different values of relative dipolar interaction
strength $\epsilon_{\mathrm{dd}}=0$ (solid), $\epsilon_{\mathrm{dd}}=0.6$
(dotted) and gas parameter $\gamma=0.01$ (red), $\gamma=0.20$ (blue).}
\end{figure}

\subsection{Superfluid depletion}

\begin{figure}[t]
\begin{centering}
\includegraphics[scale=0.45]{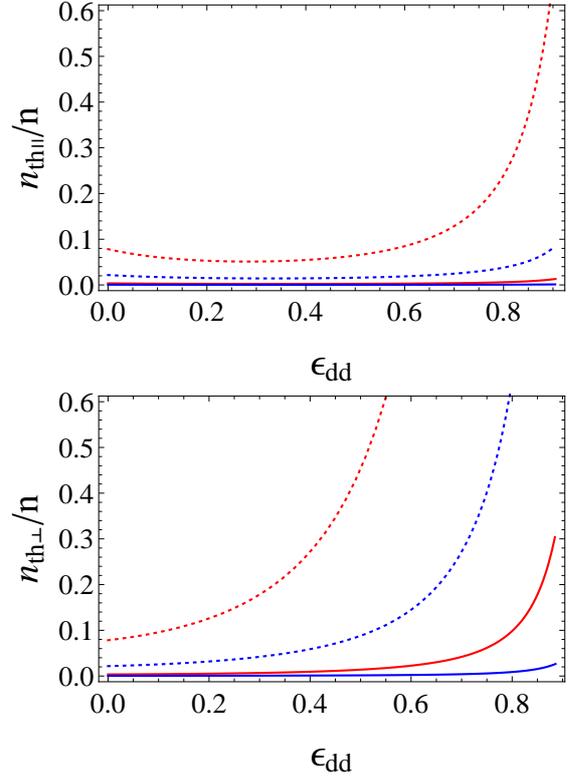}
\par\end{centering}
\protect\caption{\label{fig:Superfluid-thermal-fractional1}(color online). Superfluid thermal fractional
depletions $n_{\mathrm{th\Vert}}/n$ and $n_{\mathrm{th\bot}}/n$
versus relative dipolar interaction strength $\epsilon_{\mathrm{dd}}$
for different values of relative temperature $t=0.2$ (solid), $t=0.6$
(dotted) and gas parameter $\gamma=0.01$ (red), $\gamma=0.20$ (blue).}
\end{figure}

Now we calculate the superfluid depletion for the dipolar Bose gas
which is due to thermal fluctuations. In the thermodynamic limit of equation (\ref{eq:40})
the components parallel and perpendicular to the dipole polarization
direction have to be evaluated separately. Parallel to the dipoles the integral reads
in dimensionless form 
\begin{alignat}{1}
\frac{n_{\mathrm{th\Vert}}}{n}=\frac{\gamma^{-\frac{5}{6}}t^{4}}{8\pi^{\frac{1}{2}}\left(\zeta(\frac{3}{2})\right)^{\frac{8}{3}}}I_{\Vert}(\gamma,\epsilon_{\mathrm{dd}},t) & ,\label{eq:53}
\end{alignat}
where the remaining integral $I_{\Vert}(\gamma,\epsilon_{\mathrm{dd}},t)$
is written explicitly as
\begin{alignat}{1}
I_{\Vert}(\gamma,\epsilon_{\mathrm{dd}},t)=\int_{0}^{\infty}dx\int_{0}^{\pi}d\theta\frac{x^{4}\sin\theta\cos^{2}\theta e^{\sqrt{x^{2}+\frac{\alpha x^{4}}{16\varTheta^{2}}}}}{\varTheta^{\frac{5}{2}}\left(e^{\sqrt{x^{2}+\frac{\alpha x^{4}}{16\varTheta^{2}}}}-1\right)^{2}} & .
\end{alignat}
In the direction perpendicular to the dipoles the superfluid depletion
reads in dimensionless form 
\begin{alignat}{1}
\frac{n_{\mathrm{th\bot}}}{n}=\frac{\gamma^{-\frac{5}{6}}t^{4}}{8\pi^{\frac{1}{2}}\left(\zeta(\frac{3}{2})\right)^{\frac{8}{3}}}I_{\bot}(\gamma,\epsilon_{\mathrm{dd}},t) & ,\label{eq:55}
\end{alignat}
where the remaining integral $I_{\bot}(\gamma,\epsilon_{\mathrm{dd}},t)$
is written explicitly as 
\begin{alignat}{1}
I_{\bot}(\gamma,\epsilon_{\mathrm{dd}},t)=\int_{0}^{\infty}dx\int_{0}^{\pi}d\theta\frac{x^{4}\sin^{3}\theta e^{\sqrt{x^{2}+\frac{\alpha x^{4}}{16\varTheta^{2}}}}}{2\varTheta^{\frac{5}{2}}\left(e^{\sqrt{x^{2}+\frac{\alpha x^{4}}{16\varTheta^{2}}}}-1\right)^{2}} & .
\end{alignat}
Fig.~\ref{fig:The-superfluid-thermal} shows how the superfluid thermal
fractional depletions parallel and perpendicular to the polarized
dipoles depend on the respective values of the parameters $\gamma$,
$\epsilon_{\mathrm{dd}}$, and $t$. The perpendicular superfluid
depletion $n_{\mathrm{th\bot}}$ behaves similar to the condensate
thermal depletion $n_{\mathrm{th}}$, i.e. $n_{\mathrm{th\bot}}$
is enhanced for increasing $t$ and $\epsilon_{\mathrm{dd}}$, whereas
it is reduced for increasing $\gamma$. Also for $n_{\mathrm{th\Vert}}$
we observe that it depends on $t$ and $\gamma$, like $n_{\mathrm{th}}$
and $n_{\mathrm{th\bot}}$, i.e. it increases with $t$ but decreases
with $\gamma$. But, surprisingly, it turns out that $n_{\mathrm{th\Vert}}$
has a nontrivial dependence on $\epsilon_{\mathrm{dd}}$, where we
have first a decrease and then an increase for increasing $\epsilon_{\mathrm{dd}}$.
Fig.~\ref{fig:Superfluid-thermal-fractional1} highlights in more
detail how the superfluid thermal fractional depletions changes versus
relative dipolar interaction strength $\epsilon_{\mathrm{dd}}$.

\begin{figure}[t]
\centering{}\includegraphics[scale=0.8]{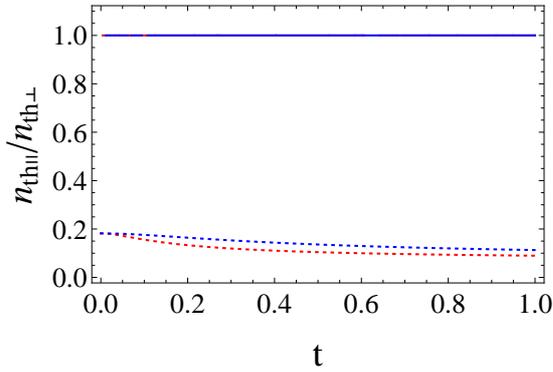}\protect\caption{\label{fig:The-thermal-superfluid1}(color online). Ratios of thermal superfluid depletions
$n_{\mathrm{th}\parallel}/n_{\mathrm{th}\perp}$ versus relative temperature
$t=T/T_{c}^{0}$ for different values of relative dipolar strength $\epsilon_{\mathrm{dd}}=0$
(solid), $\epsilon_{\mathrm{dd}}=0.6$ (dotted) and gas parameter
$\gamma=0.01$ (red), $\gamma=0.20$ (blue).}
\end{figure}

\begin{figure}[t]
\centering{}\includegraphics[scale=0.8]{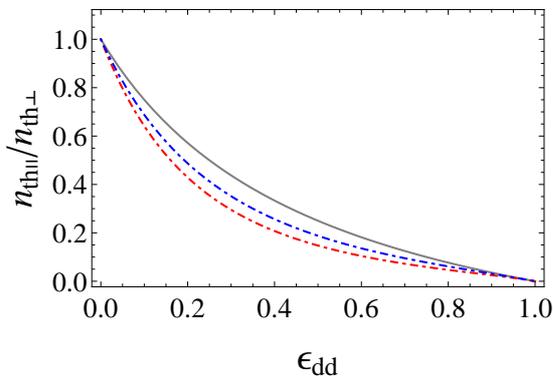}\protect\caption{\label{thermalsuper}(color online). Ratios of thermal superfluid depletions $n_{\mathrm{th}\parallel}/n_{\mathrm{th}\perp}$
versus relative dipolar interaction strength $\epsilon_{\mathrm{dd}}$
for different values of the gas parameter $\gamma=0.01$ (red), $\gamma=0.20$
(blue) and relative temperature $t=0$ (solid-gray), $t=0.5$ (dotted-dashed)
with $t=0$ limit from (\ref{eq:49}), (\ref{eq:50-1}).}
\end{figure}

Note that both $n_{\mathrm{th\bot}}$ and $n_{\mathrm{th\Vert}}$
coincide for vanishing dipolar interaction as follows from Fig.~\ref{fig:The-thermal-superfluid1}
which shows the thermal superfluid depletion ratio $n_{\mathrm{th}\parallel}/n_{\mathrm{th}\perp}$
versus the relative temperature $t$. When the dipolar interaction
is present we notice that the ratio $n_{\mathrm{th}\parallel}/n_{\mathrm{th}\perp}$
reveals only a tiny $t$- and $\gamma$-dependence which is decreasing
with $t$ and increasing with $\gamma$, respectively. In Fig.~\ref{thermalsuper}
we plot the thermal superfluid depletions ratio $n_{\mathrm{th}\parallel}/n_{\mathrm{th}\perp}$
versus the relative dipolar strength $\epsilon_{\mathrm{dd}}$. It
shows that both $n_{\mathrm{th\bot}}$ and $n_{\mathrm{th\Vert}}$
coincide for vanishing dipolar interaction, and the ratio $n_{\mathrm{th}\parallel}/n_{\mathrm{th}\perp}$
decreases for increasing values of the relative dipolar strength $\epsilon_{\mathrm{dd}}$.

In the vicinity of the zero temperature but far below critical temperature
we can approximate the superfluid depletion in (\ref{eq:53}) analytically.
We find that the superfluid depletion parallel to the dipoles reads
\begin{alignat}{1}
\frac{n_{\mathrm{th\Vert}}}{n}=\frac{\pi^{\frac{7}{2}}\gamma^{-\frac{5}{6}}t^{4}}{15\left(\zeta(\frac{3}{2})\right)^{\frac{8}{3}}}\, J_{-\frac{5}{2}}(\epsilon_{\mathrm{dd}})+.. & .\label{eq:49}
\end{alignat}
which has a non-monotonic behavior versus $\epsilon_{\mathrm{dd}}$
due to the function $J_{-\frac{5}{2}}(\epsilon_{\mathrm{dd}})$, which
is plotted in Fig.~\ref{fig:The-functionJ} where we have a decrease
at first and then an increase for increasing $\epsilon_{\mathrm{dd}}$.
Furthermore, the superfluid depletion perpendicular to the dipoles
reads
\begin{alignat}{1}
\frac{n_{\mathrm{th\bot}}}{n}=\frac{\pi^{\frac{7}{2}}\gamma^{-\frac{5}{6}}t^{4}}{30\left(\zeta(\frac{3}{2})\right)^{\frac{8}{3}}}\left[Q_{-\frac{5}{2}}(\epsilon_{\mathrm{dd}})-J_{-\frac{5}{2}}(\epsilon_{\mathrm{dd}})\right]+.. & .\label{eq:50-1}
\end{alignat}
so that $n_{\mathrm{th\bot}}$ is enhanced for increasing $\epsilon_{\mathrm{dd}}$
due to the fact that the function $Q_{-\frac{5}{2}}(\epsilon_{\mathrm{dd}})$
shown in Fig.~\ref{fig:The-dipolar-enhancement} increases faster
than the function $J_{-\frac{5}{2}}(\epsilon_{\mathrm{dd}})$ shown
in Fig.~\ref{fig:The-functionJ}. A comparison between the two different
superfluid depletions shows that both $n_{\mathrm{th\bot}}$ and $n_{\mathrm{th\Vert}}$
coincide for vanishing dipolar interaction, and reproduce the isotropic
contact interaction result \cite{landau-1} due to $Q_{\alpha}(0)=1$
and $J_{\alpha}(0)=1/3$. The ratio $n_{\mathrm{th}\parallel}/n_{\mathrm{th}\perp}$
from (\ref{eq:49}), (\ref{eq:50-1}) decreases for increasing values
of the relative dipolar strength $\epsilon_{\mathrm{dd}}$ as is shown
in gray color in Fig.~\ref{thermalsuper}

\begin{figure}[t]
\begin{centering}
\includegraphics[scale=0.45]{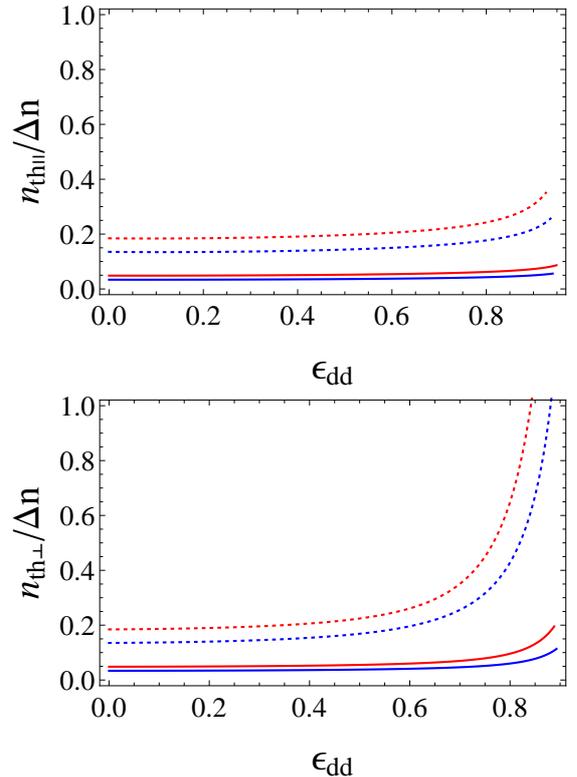}
\par\end{centering}
\protect\caption{\label{fig:Ratios-of-superfluid1}(color online). Ratios of superfluid and condensate
depletions $n_{\mathrm{th}\parallel}/\Delta n$ (upper) and $n_{\mathrm{th}\perp}/\Delta n$
(lower) versus relative dipolar strength $\epsilon_{\mathrm{dd}}$
for different values of relative temperature $t=0.3$ (solid), $t=0.6$
(dotted) and gas parameter $\gamma=0.01$ (red), $\gamma=0.11$ (blue)
in the clean case $R_{0}=0$.}
\end{figure}

\begin{figure}[t]
\begin{centering}
\includegraphics[scale=0.45]{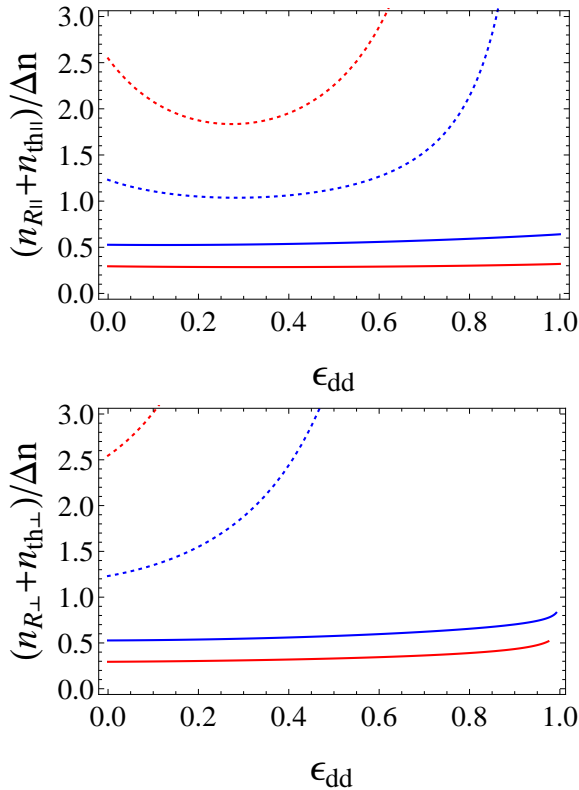}
\par\end{centering}
\protect\caption{\label{fig:Superfluid-depletions-ratios}(color online). Ratios of superfluid depletions
$(n_{\mathrm{th}\parallel}+n_{\mathrm{R}\parallel})/\Delta n$ (upper)
and $(n_{\mathrm{th}\perp}+n_{\mathrm{R}\perp})/\Delta n$ (lower)
versus relative dipolar strength $\epsilon_{\mathrm{dd}}$ for different
values of relative temperature $t=0$ (solid), $t=0.5$ (dotted) and
gas parameter $\gamma=0.01$ (red), $\gamma=0.08$ (blue) in the dirty
case $R_{0}=\frac{2\hbar^{4}\pi^{\frac{3}{2}}n^{\frac{1}{3}}}{5m^{2}}$.}
\end{figure}

\subsection{Superfluid Ratios}

Finally after having discussed the different origins of the superfluid
depletions for the dipolar Bose gas, we discuss now the ratios of
the total superfluid depletions over the total condensate depletion
$\Delta n$ for different values of the respective parameters $\gamma$,
$\epsilon_{\mathrm{dd}}$, and $t$. We consider both the clean case
where the disorder strength $R_{0}=0$ and in the dirty case with
$R_{0}=\frac{2\hbar^{4}\pi^{\frac{3}{2}}n^{\frac{1}{3}}}{5m^{2}}$.
Fig.~\ref{fig:Ratios-of-superfluid1} shows the clean case with the
ratios $n_{\mathrm{th}\parallel}/\Delta n$ (upper) and $n_{\mathrm{th}\perp}/\Delta n$
(lower) versus relative dipolar strength $\epsilon_{\mathrm{dd}}$.
For increasing $\epsilon_{\mathrm{dd}}$ and $t$, both $n_{\mathrm{th}\parallel}/\Delta n$
and $n_{\mathrm{th}\perp}/\Delta n$ are enhanced, but increasing
$\gamma$ decreases both $n_{\mathrm{th}\parallel}/\Delta n$ and
$n_{\mathrm{th}\perp}/\Delta n$.

Fig.~\ref{fig:Superfluid-depletions-ratios} shows in the dirty case
the ratios $(n_{\mathrm{th}\parallel}+n_{\mathrm{R}\parallel})/\Delta n$
(upper) and $(n_{\mathrm{th}\perp}+n_{\mathrm{R}\perp})/\Delta n$
(lower) versus relative dipolar strength $\epsilon_{\mathrm{dd}}$.
For increasing $\epsilon_{\mathrm{dd}}$ we will get enhancement for
$(n_{\mathrm{th}\perp}+n_{\mathrm{R}\perp})/\Delta n$, regardless
of the temperature $t$, while for $(n_{\mathrm{th}\parallel}+n_{\mathrm{R}\parallel})/\Delta n$
we will get enhancement for small values of $t$ and then upon increasing
$t$, we observe a decrease up to $\epsilon_{\mathrm{dd}}=0.3$ and
then an increase, which shows that the disorder changes the situation
a lot compared to the clean case in Fig.~\ref{fig:Ratios-of-superfluid1}.
Increasing $\gamma$ will cause a more complex interplay on both $(n_{\mathrm{th}\parallel}+n_{\mathrm{R}\parallel})/\Delta n$
and $(n_{\mathrm{th}\perp}+n_{\mathrm{R}\perp})/\Delta n$, as is
seen in Fig.~\ref{fig:Superfluid-depletions-ratios}.

\section{Conclusion}

In this paper we have investigated the dipolar dirty boson problem,
where both the short-range isotropic contact interaction as well as
the anisotropic long-range dipole-dipole interaction are present within
a Bogoliubov theory at low temperatures below the critical temperature.
In the zero-temperature limit, we discussed analytically the nontrivial
results for both the condensate and the superfluid depletions as well as
the interplay between the long-range anisotropic dipolar interaction
and the external random potential which yields an anisotropic superfluidity
\cite{pelster1,pelster3}. At finite temperature, we obtained, due
to the long-range anisotropic dipole-dipole interaction, different
analytical and numerical results for condensate and the superfluid
depletions, where the latter depends sensitively on whether the flow
direction is parallel or perpendicular to the dipolar polarization
axis. It still remains to investigate how the Josephson relation between condensate and superfluid densities reads
for a disordered dipolar superfluid \cite{cord}. \\[4mm]

\section{Acknowledgment}

We both acknowledge fruitful discussions with T. Checinski and A.
R. P. Lima. M. Ghabour thanks his family for financial support during
this work. A. Pelster thanks for the support from the German Research
Foundation (DFG) via the Collaborative Research Center SFB /TR49 Condensed
Matter Systems with Variable Many-Body Interactions.

\end{document}